\pdfoutput=1

\documentclass[11pt]{article}
\usepackage{enumitem}
\usepackage[final]{acl}
\usepackage{times}
\usepackage{latexsym}

\usepackage[T1]{fontenc}

\usepackage[utf8]{inputenc}

\usepackage{microtype}

\usepackage{inconsolata}

\usepackage{graphicx}
\usepackage{xcolor}
\usepackage{tcolorbox}
\usepackage{booktabs}    
\usepackage{tabularx}    
\usepackage{multirow}    
\usepackage{ragged2e}    
\usepackage{caption}     
\usepackage{inconsolata}
\usepackage[flushmargin]{footmisc}
\makeatletter
\def\thanks#1{\protected@xdef\@thanks{\@thanks\protect\footnotetext{#1}}}
\makeatother
\title{EMO-RL: Emotion-Rule-Based Reinforcement Learning Enhanced Audio-Language Model for Generalized Speech Emotion Recognition}

\author{
 \textbf{Pengcheng Li\textsuperscript{1,2}\textsuperscript{$\dagger$}},
 \textbf{Botao Zhao\textsuperscript{1}\textsuperscript{$\dagger$}},
 \textbf{Zuheng Kang\textsuperscript{1}},
\\
 \textbf{Junqing Peng\textsuperscript{1}},
 \textbf{Xiaoyang Qu\textsuperscript{1}},
 \textbf{Yayun He\textsuperscript{1}},
 \textbf{Jianzong Wang\textsuperscript{1}\textsuperscript{*}}
\\
 \textsuperscript{1}Ping An Technology (Shenzhen) Co., Ltd., Shenzhen, China,
\\
 \textsuperscript{2}Tsinghua Shenzhen International Graduate School, Tsinghua University, Shenzhen, China,
\\
 \thanks{
 \textsuperscript{$\dagger$} These authors contributed equally to this work. \\
 \textsuperscript{*} Corresponding author: \href{jzwang@188.com}{jzwang@188.com}.
 }
}
\usepackage{amssymb, amsmath, color, framed}
\usepackage{amsmath}
\usepackage{indentfirst} 
\usepackage{multirow}
\begin{document}
\maketitle
\begin{abstract}
Although Large Audio-Language Models (LALMs) have exhibited outstanding performance in auditory understanding, their performance in affective computing scenarios, particularly in emotion recognition, reasoning, and subtle sentiment differentiation, remains suboptimal.
Recent advances in Reinforcement Learning (RL) have shown promise in improving LALMs' reasoning abilities. 
However, two critical challenges hinder the direct application of RL techniques to Speech Emotion Recognition (SER) tasks: (1) convergence instability caused by ambiguous emotional boundaries and (2) limited reasoning ability when using relatively small models (e.g., 7B-parameter architectures).
To overcome these limitations, we introduce EMO-RL, a novel framework incorporating reinforcement learning with two key innovations: Emotion Similarity-Weighted Reward (ESWR) and Explicit Structured Reasoning (ESR).
Built upon pretrained LALMs, our method employs group-relative policy optimization with emotion constraints.
Comprehensive experiments demonstrate that our EMO-RL training strategies can significantly enhance the emotional reasoning capabilities of LALMs, attaining state-of-the-art results on both the MELD and IEMOCAP datasets, and cross-dataset experiments prove the strong superiority of generalization.
\end{abstract}

\section{Introduction}
Speech Emotion Recognition (SER) is a significant research direction in the field of affective computing, aiming to map speech signals to corresponding emotional labels through computational analysis. It plays an important role in various applications, including intelligent customer service~\citep{li2021speech}, mental health assessment~\citep{madanian2022automatic}, and human-computer interaction~\citep{alsabhan2023human}. 

\begin{figure}[h]
  \centering
  \includegraphics[width=0.49\textwidth]{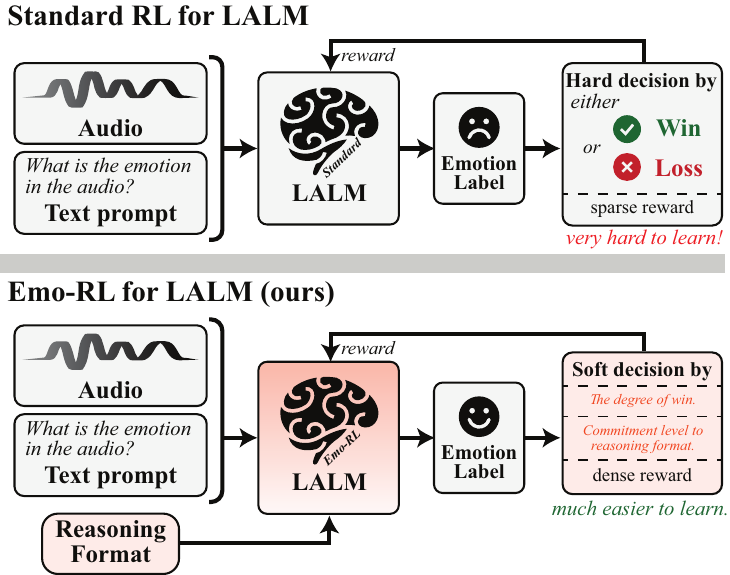}
  \caption{The key ideas of our proposed Emo-RL. Compared with the standard RL, Emo-RL exploited emotion similarity-weighted reward and the explicit structured reasoning to improve the emotion recognition performance of LALM.}
  \label{fig:motivation}
\end{figure}

In SER task, prior studies predominantly rely on pre-trained speech models or perform fine-tuning on affective corpora to derive emotional representations, then train a classification head to implement emotion classification~\citep{li2024multiscaletemporaltransformerspeech, chen2023dst}. However, the emotional representations extracted by these models can only capture the acoustic expressions of emotions, but lack a collaborative analysis of text semantics. These models have very limited generalization capability and lack explainability.

With the development of multi-modal large models, many powerful Large Audio-Language Models (LALMs) have emerged~\citep{kong2024audio}, among which Qwen2-Audio~\citep{chu2024qwen2} is a representative example. 
It can follow user instructions to perform many downstream tasks, such as speech recognition, transcription, sound classification, and more. Although Qwen2-Audio demonstrates strong speech understanding capabilities, its performance on SER tasks remains limited. 
This limitation stems from its tendency to rely on shallow associations rather than multi-step reasoning that integrates textual semantics and auditory features across modalities, resulting in its limited accuracy and generalization on SER.
Speech emotion recognition inherently constitutes a cognitive reasoning process that necessitates comprehensive analysis from multiple perspectives through step-by-step reasoning.
When humans recognize emotions in speech, they often understand the specific content and keywords of speech and integrate this understanding based on acoustic features (such as pitch, voice quality, speech rate). 
For example, when someone says "fed up" with rapid speech, high volume, and sharp intonation, anger can be inferred. 

These reasoning steps are beyond the capability of traditional audio feature extraction and classification head frameworks. 
Extensive research has shown that reinforcement learning can enhance the reasoning capabilities of LLM~\citep{guo2025deepseek,team2025kimi}. 
The effective deployment of reinforcement learning (RL) in SER encounters two fundamental limitations: convergence reliability issues primarily arising from ill-defined inter-class affective boundaries that induce gradient conflict during policy updates, compounded by insufficient affective reasoning capacity in under-parameterized architectures (e.g., 7B-parameter configurations).

To address these challenges, we adopt a psychological perspective by transforming the original right/wrong classification problem into a regression problem that accommodates varying degrees of correctness and error, through the introduction of an emotion-state-transition matrix (As illustrated in Figure~\ref{fig:motivation}).
 We implement an Emotion Similarity-Weighted Reward (ESWR) mechanism that progressively guides the policy model from simpler to more complex tasks. This approach initially teaches the model to distinguish between basic positive and negative emotions before advancing to finer-grained emotional distinctions.
 To further enhance the model's emotional reasoning capabilities, we incorporate Explicit Structured Reasoning (ESR) strategies  during RL training. These strategies provide the model with guiding clues to help it more effectively differentiate between emotions, thereby improving its overall reasoning ability in SER tasks.

Based on the ESWR and ESR, we exploited our emotion-rule-based RL method to fine-tune the LALM, and the contributions of this paper are summarized as follows:
\begin{itemize} [leftmargin=10pt]
\item We propose a SER pipeline via RL fine-tuning of a large audio and language model.
\item We introduce Emotion-rule based RL to improve the emotion recognition ability of LALM, leveraging emotion similarity-weighted rewards and explicit structured reasoning strategies.
\item Extensive experiments demonstrate that the proposed approach exhibits strong generalizability and achieves state-of-the-art performance.
\end{itemize}

\section{Related Works}
\subsection{Generalized Speech Emotion Recognition}
For SER, traditional approaches have focused on designing novel network architectures~\citep{zou2022speech, li2024multiscaletemporaltransformerspeech} based on classical neural networks.
With the advancement of self-supervised learning, researchers have increasingly utilized pre-trained audio models like WavLM~\citep{chen2022wavlm}, Emotion2vec~\citep{ma2023emotion2vec}, HuBERT~\citep{hsu2021}, and Whisper~\citep{radford2023robust} to extract speech features or fine-tune these models on speech emotion datasets to obtain emotion-specific features~\citep{morais2022speech, chen2023exploring}. 
Subsequently, a linear classification head is trained to perform emotion classification. 
These models have significantly enhanced speech emotion perception capabilities~\citep{li2023exploration}. 
For instance, the Vesper model~\citep{chen2024vesper}, obtained by distilling the WavLM-large model with emotion data, has achieved promising results in SER tasks.  
However, the generalization capabilities of these models remain limited, and they lack collaborative analysis of text semantics and explainability.

\subsection{Large Audio-Language Models}
Recent progress in multimodal large-scale language modeling has led to the emergence of numerous LALMs, such as Audio Flamingo~\citep{kong2024audio}, Qwen2-Audio~\citep{chu2024qwen2},and SALMONN~\citep{tang2023salmonn}, which have demonstrated strong audio understanding capabilities, with Qwen2-Audio even outperforming previous methods across the vast majority of audio-focused evaluation benchmark.
These models typically comprise three main components: an Audio Encoder, a Large Language Model, and a modality connector that bridges them. 
These models are capable of directly processing cross-modal inputs, including audio (such as speech, environmental sounds, and music) and text prompts, and can generate the corresponding textual output. They are able to follow user instructions to perform a variety of downstream tasks, such as  transcription, SER, and sound classification~\citep{wang2025enabling, waheed2024speech}.
However, current LALM training mainly focuses on perception and basic QA tasks, lacking explicit multi-step reasoning. Thus, the potential of LALMs like Qwen2-Audio in complex audio reasoning tasks such as SER remains untapped. Enhancing their reasoning abilities in these advanced tasks is crucial.

\subsection{Reinforcement Learning and Reasoning}
Reinforcement learning (RL) plays a crucial role in advancing the reasoning abilities of LLMs and MLLMs.
RLHF employs proximal policy optimization (PPO)\citep{schulman2017proximal} alongside a trained reward mechanism to align LLMs with human preferences.
Direct Preference Optimization (DPO)\citep{rafailov2023direct} bypasses reward modeling by learning from preference data directly, whereas Rejection Sampling Fine-tuning (RFT)\citep{yuan2023scaling} strengthens reasoning through curated self-produced reasoning chains.
Group Relative Policy Optimization (GRPO)\citep{shao2024deepseekmath} refines PPO by eliminating the critic component and utilizing group-level baseline averaging for advantage computation, achieving enhanced LLM reasoning with reduced computational overhead.
The Hybrid GRPO variant~\citep{sane2025hybrid} integrates GRPO's sampling mechanism with a trained value estimator, improving training stability and data utilization efficiency.
Contemporary research demonstrates that Chain-of-Thought (COT) combined with RL substantially elevates LALM reasoning capabilities. Audio-CoT~\citep{ma2025audio} pioneered COT integration in LALMs, though gains were modest without model parameter optimization.
Audio-Reasoner~\citep{xie2025audio} developed CoTA, an extensive synthetic corpus containing millions of question-answer instances with detailed reasoning trajectories, markedly advancing extended-context reasoning abilities.
Xiaomi's implementation utilized GRPO optimization on the Qwen2-Audio-7B architecture for audio question-answering applications~\citep{li2025reinforcement}, achieving notable improvements in reasoning precision.
SARI~\citep{wen2025sari} additionally combines systematic reasoning frameworks with progressive reinforcement training curricula, establishing new benchmarks on MMAU and MMSU evaluations.
Reward-based optimization frameworks have proven effective in boosting reasoning precision, demonstrating that reinforcement-driven strategies can maximize learning efficiency from constrained training datasets. Nevertheless, existing RL methodologies remain overly broad for specialized speech emotion applications. Consequently, developing emotion-rule-guided reinforcement strategies specifically tailored for SER tasks becomes essential.

\section{Methodology}
\begin{figure*}[h]
  \centering
  \includegraphics[width=0.99\linewidth]{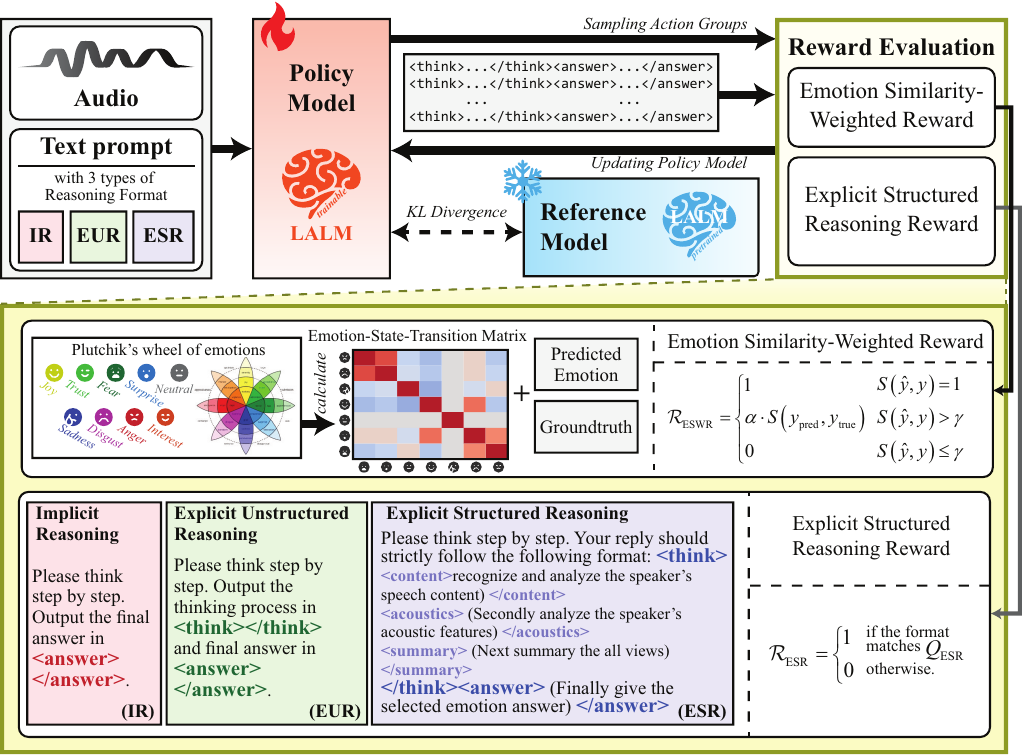}
  \caption{The overview of the Emo-RL for LALM to improve the generalized speech emotion recognition. Building on GRPO, we enhance emotion recognition via two improvements. First, we create an emotion-state-transition matrix from Plutchik's wheel of emotions~\cite{plutchik1982psychoevolutionary}, allowing the policy model to receive rewards for predicting similar emotions. Second, we introduce explicit structured reasoning to directly input human emotion recognition priors into the model.}
  \label{fig:main_figure}
  \vspace{-0.5cm}
\end{figure*}

\subsection{Problem Definition}
We use emotional audio question answering in Qwen2-Audio-7B-Instruct to implement SER.
SER process through LALMs constitutes a parametric mapping process where: given a speech signal $x$ and structured textual query $Q$ containing multiple-choice options, their temporal-contextual concatenation forms the input prompt $p=[x;Q]$. The LALM, $\pi_{\boldsymbol{\theta}}$, then generates emotion prediction $\hat{y}$ through cross-modal understanding, formally expressed as:
\begin{equation}
    \pi_{\boldsymbol{\theta}}(x; Q)\to a \to \hat{y},
    \label{forward}
\end{equation}
where $S$ is a speech audio with a sampling rate of 16kHz, $Q$ is the textual question prompt, and $a$ is the generated response of LALM, including thinking and reasoning contents and the final selected answer, and $\hat{y}$ denotes the predicted emotion label. 

This study aim to address two core challenges in SER through LALMs: (1) Enhancing the predictive accuracy of $f_\theta$ via reinforcement learning using the training dataset $\mathcal{D}=\{(x_{i}, y_{i})\}^{N}_{i}$, where the N means the sample number. (2) Discovering optimal prompt formulations $Q$ to improve the inference performance. Considering the parameter space of is infinite, we defined three experimentally validated designs as the Q space, $\mathbf{Q}$, including implicit reasoning $Q_{\mathrm{IR}}$, explicit unstructured reasoning $Q_{\mathrm{EUR}}$, explicit structured reasoning $Q_{\mathrm{ESR}}$. Therefore, the target of this study could be defined as:
\begin{equation}
    \theta, Q = \arg \max_{\theta, Q \in \mathbf{Q}} (\mathcal{R}(\pi_{\boldsymbol{\theta}}(X;Q), Y)), 
    \label{object}
\end{equation}
where the $R$ denotes the reward function.


In detail, we explore three reasoning strategies in
EMO-RL training to evaluate the impact of reasoning patterns. We detail three patterns below:  
\begin{itemize}[leftmargin=10pt]
    \item Implicit Reasoning, $Q_{\mathrm{IR}}$: The foundational configuration involves training the system to directly produce conclusive responses, bypassing any obligation to articulate underlying cognitive mechanisms or intermediate analytical steps.
    
    \item Explicit Unstructured Reasoning, $Q_{\mathrm{EUR}}$: This approach facilitates organic and unconstrained thought expression by employing prompting techniques that eschew rigid organizational templates or prescribed divisions. While permitting flexible formulation, the framework necessitates generation of logically consistent interpretations culminating in unambiguous determinations.
    
    \item Explicit Structured Reasoning, $Q_{\mathrm{ESR}}$: The methodology enforces systematic generation of transparently organized cognitive pathways. Implementation requires adherence to a bifurcated analytical framework encompassing textual dimensions (verbatim transcriptions, pivotal terminology) and prosodic characteristics (intonation patterns, temporal cadence, acoustic intensity, vocal texture). Through synthesis of these dual information streams, the system derives its conclusive assessment.
\end{itemize}

\subsection{Emotion-rule based RL framework}
We built our Emo-RL based on the GRPO framework for its efficiency and scalability. Unlike proximal policy optimization, which requires a computationally expensive value network, GRPO calculates relative advantages by comparing rewards within a group of sampled actions, reducing computational overhead and simplifying optimization. This makes GRPO particularly suitable for speech reasoning tasks. Similar to GRPO, the Emo-RL also has three main steps, sampling action groups, reward evaluation, and updating policy network with relative advantage and KL divergence (As shown in Figure~\ref{fig:main_figure}).

\textbf{Sampling Action Groups} For each input state $s=(x,Q)$ , where $x$ is the speech encoding of the input audio and $Q$ the textual encoding of the question, GRPO samples a group of actions (the generated response of LALM), $\{a_{1},a_{2},\ldots,a_{G}\}$, from the current policy $\pi_{\boldsymbol{\theta}}$. The sampling process is:  
\begin{equation}
    a_{i}\sim\pi_{\theta}(a\mid x,Q),\quad{\mathrm{for}}\quad i=1,2,\ldots,G.
\end{equation}
This strategy ensures diverse responses, promoting exploration and preventing premature convergence.  

\textbf{Reward Evaluation}. In our reinforcement learning framework, each sampled action $a_{i}$ is assigned a reward $\mathcal{R}(a_{i})$ based on verifiable criteria, resulting in a reward set $\{r_{1},r_{2},\dots,r_{G}\}$. For emotional speech reasoning tasks, the reward function $\mathcal{R}(a_{i})$ combines two components: the reasoning format reward $\mathcal{R}_{\mathrm{format}}(a_{i})$ and emotion accuracy reward $\mathcal{R}_{\mathrm{acc}}(a_{i})$. The format reward ensures that the responses adhere to a structured format, thereby guiding the reasoning strategy of the policy network, $\pi_{\theta}$. The accuracy reward evaluates the correctness of the action $a_{i}$, providing feedback to $\pi_{\theta}$ on the extent to which the answer aligns with the correct response. The overall reward function is:  
\begin{equation}
    \mathcal{R}(a_{i})=\mathcal{R}_{\mathrm{format}}(a_{i})+\mathcal{R}_{\mathrm{acc}}(a_{i}).
\end{equation}

\textbf{Updating Policy Network with Relative Advantage and KL divergence}. The $\pi_{\theta}$ is optimized by the Relative Advantage of rewards and KL divergence between $\pi_{\theta}$ and reference model $\pi_{\mathrm{ref}}$. Firstly, policy Rewards are normalized within the sampled group to compute relative advantages $\{A_{1},A_{2},\dotsc,A_{G}\}$, defined as:  
\begin{equation}
    A_{i}={\frac{r_{i}-\operatorname*{mean}\{r_{1},r_{2},\ldots,r_{G}\}}{\operatorname{std}\{r_{1},r_{2},\ldots,r_{G}\}}}.
\end{equation}
Based on these advantages, the policy is updated to reinforce actions with positive advantages and reduce the probability of less effective ones. To ensure stable RL learning, $\pi_{\theta}$ updates are further constrained by minimizing the KL divergence between the updated and reference models. 

\subsection{Rewards Mechanism Design}
The EMO-RL framework implements dual reward mechanisms synergistically combining structural compliance enforcement and affective alignment optimization. 
Specifically, domain-specific response schemata are enforced through regular-expression pattern matching that validates three distinct reasoning pattern compliance rates ($Q_{\text{IR}}, Q_{\text{EUR}}, Q_{\text{ESR}}$), systematically enhancing explainability via cognitive transparency in decision pathways. 
Complementarily, the emotion similarity-weighted reward employs an emotion-state-transition matrix constructed through Plutchik's wheel of emotions~\cite{plutchik1982psychoevolutionary}, generating dense reward signals that precisely guide policy gradients through convex optimization landscapes. 
\subsubsection{Reasoning Format Reward}

This component ensures adherence to specific response formats across different reasoning strategies by implementing tailored format reward. We define three distinct reasoning format functions, including Implicit Reasoning (IR), Explicit Unstructured Reasoning (EUR), and Explicit Structured Reasoning (ESR), each requiring different format constraints.

For IR, which targets only answer generation, the reward is granted only when the final answer is both correct and correctly delimited by \texttt{<answer>} tags, as shown in Figure \ref{fig:main_figure}.
For EUR, which require explicit reasoning display, the format reward is granted when the response contains reasoning within \texttt{<think>} tags and the final answer within \texttt{<answer>} tags. ESR is similar to EUR, but with four additional format constraint tags.

The format reward function of ESR is as follows, and all format reward follow this rule. Each format reward function employs binary scoring based on regex pattern matching, where strict adherence to the specified format yields a reward score of 1, while any deviation results in a score of 0. This ensures consistent formatting across different reasoning strategies while maintaining the flexibility to accommodate strategy-specific requirements.

\begin{equation}
    \mathcal{R}_{\mathrm{ESR}}=\begin{cases}1,&\text{if the format matches $Q_{\text{ESR}}$}   \\ 0, &\mathrm{otherwise.}\end{cases}
\end{equation}
\subsubsection{Emotion Accuracy Reward}
The conventional approach uses binary classification rewards (BCR), allocating a score of 1 to fully accurate responses and 0 to all others. However, this has limitations when applied to emotions, as it ignores the relationships between different emotion types. Emotions are inherently continuous and complex. Drawing from psychological emotion dimension theories~\cite{plutchik1982psychoevolutionary} and other psychological knowledge, we comprehensively consider emotion valence (the positive or negative of emotions) and arousal (the intensity or activation level of emotions). We have meticulously designed an emotion-state-transition matrix $S \in \mathbb{R}^{C \times C}$ ($C$ denote emotion categories) as:

\begin{equation}
\begin{footnotesize}
S_{i,j}=\left\{ \begin{matrix}
	\frac{1}{2},&		\text{if $y_i$ or $y_i$=`$\mathrm{neutral}$'}\\
	\frac{1}{2}\left( \cos \left( \text{Pl}\left( y_i,y_j \right) \right) +1 \right) ,&		\mathrm{otherwise}\\
\end{matrix} \right.
\end{footnotesize}
\end{equation}

Here, the $y_{i}$ and $y_{j}$ denotes two emotion types, and $\mathrm{Pl}(\cdot, \cdot)$ means the angles from each other on the Plutchik's wheel of emotions.
Based on the $S$, we have the Emotion Similarity-Weighted Reward function, formulated as:
\begin{equation}
    \mathcal{R}_{\mathrm{ESWR}} =
\begin{cases}
1 & S(\hat{y}, y) = 1 \\
\alpha \cdot S(\hat{y}, y) & S(\hat{y}, y)>\gamma \\
0 & S(\hat{y}, y)<=\gamma 
\end{cases}
\end{equation}
where $\alpha$ is the partial matching coefficient, dynamically adjusting from 1 to 0 during training, and $\gamma$ is the threshold of the contradictory emotion, which was set as 0.7 in this paper.





\section{Experiment}
\subsection{Dataset}
 We evaluated model capabilities and generalization in speech emotion recognition (SER) using four datasets: 
 \textbf{MELD}~\citep{poria2018meld} (13 708 utterances from \emph{Friends}, 7 emotions), \textbf{IEMOCAP}~\citep{busso2008iemocap} (5,531 utterances from conversations, 4 emotions), \textbf{RAVDESS}~\citep{livingstone2018ryerson} (4 800 audio-video recordings of speech and song, 8 emotions), and \textbf{SAVEE}~\citep{jackson2014surrey} (480 samples, 7 emotions).

\subsection{Implementation Details}
We use Qwen2-Audio-7B-instruct as the foundational backbone model for our experiments. 
The RL models are trained using eight NVIDIA RTX A6000 GPUs, each processing a per-device batchsize of 1 with gradient accumulation over 2 steps.
Training proceeds for 300 optimisation steps under a learning rate of $1 \times 10^{-6}$ and a softmax temperature of 1.0. 
Each reinforcement learning optimization step generates 6 responses per sample. 
SFT models are optimized with AdamW at a learning rate of  $1 \times 10^{-5}$  for five complete epochs.
The optimal iteration results are selected for final analysis.

\subsection{Baselines and Metrics}
We benchmark the proposed approach against state-of-the-art methods, which we group into three distinct categories: \textbf{W/o-LALM}, \textbf{LALM}, and \textbf{LALM-FT}. 
W/o-LALM and LALM-FT refer to models post-trained on the MELD training set, while LALM involves zero-shot inference using prompt strategies without task-specific fine-tuning.
\begin{itemize}[leftmargin=10pt]
    \item \textbf{W/o-LALM}: We selected four advanced self-supervised pre-trained audio models: HuBERT large~\cite{hsu2021},  data2vec 2.0 large~\cite{baevski2023efficient}, WavLM large~\cite{chen2022wavlm}, Whisper large v3~\cite{radford2023robust} and Emotion2vec~\cite{ma2023emotion2vec}. Features from the last Transformer layer of these frozen pre-trained models were extracted to train the downstream linear layers with a hidden dimension of 256.
    \item \textbf{LALM}: We directly use Qwen2-Audio~\cite{chu2024qwen2} for SER tasks without additional training or fine-tuning, employing two prompt patterns: direct inference and chain-of-thought inference.
    \item \textbf{LALM-FT}: We further trained Qwen2-Audio. To evaluate different training methods, we compare models trained with supervised fine-tuning (SFT), GRPO~\cite{shao2024deepseekmath}, and EMO-RL. Additionally, we assess the impact of different reasoning strategies in EMO-RL: implicit reasoning, unstructured explicit reasoning, and structured explicit reasoning.
\end{itemize}
In our evaluation, we utilize three key metrics: Unweighted Accuracy (UA), Weighted Accuracy (WA), and Macro F1 Score (F1), to assess the performance of the SER task. WA reflects the overall accuracy of the model across all emotion classes. 
UA measures the average accuracy by considering each emotion class equally, regardless of its frequency in the dataset. 
The macro-F1 score, harmonic mean of precision and recall, furnishes a balanced and class-agnostic gauge of model efficacy,  particularly in scenarios where there is an imbalance in the distribution of emotion classes.

\begin{table*}[ht]
\caption{The comparison of the main performance metrics for various methods on the MELD and IEMOCAP datasets. Results for W/o-LALM methods are cited from the Emobox benchmark~\cite{ma2024emobox} and~\cite{ma2023emotion2vec}. 
The \textbf{bold} font indicates the best results among all models.
The baseline here denotes the GRPO+IR, and the SOTA means the best results among W/o-LALM methods.}
\label{tab:performance}
\centering
\small 
\setlength{\tabcolsep}{2.4pt}               

\begin{tabularx}{\textwidth}{>{\raggedright\arraybackslash}p{1.55cm} llccccccc} 
\toprule
\multirow{2}{*}{\textbf{Model Type}} & \multirow{2}{*}{\textbf{Model}} & \multirow{2}{*}{\textbf{Method}} & \multicolumn{3}{c}{\textbf{MELD}} & \multicolumn{3}{c}{\textbf{IEMOCAP}} \\
\cmidrule(r){4-6} \cmidrule(l){7-9} 
& & & \textbf{UA($\%$)} & \textbf{WA($\%$)} & \textbf{F1($\%$)} & \textbf{UA($\%$)} & \textbf{WA($\%$)} & \textbf{F1($\%$)} \\
\midrule
\multirow{4}{*}{W/o-LALM}
& HuBERT large~\citep{hsu2021} & Classification Head & 24.13 & 46.37 & 24.99 & 67.42 & 66.69 & 67.24\\
& WavLM large~\citep{chen2022wavlm} & Classification Head & 28.18 & 49.31 & 29.11 & 69.47 & 69.07 & 69.29\\
& data2vec 2.0 large~\citep{baevski2023efficient} & Classification Head & 26.33 & 47.72 & 27.35 & 57.30 & 56.23 & 56.70\\
& Whisper large V3~\citep{radford2023robust} & Classification Head & 31.54 & 51.89 & 32.95 & 73.54 & 72.86 & 73.11\\
& Emotion2vec+ large~\citep{ma2023emotion2vec} & Classification Head & 28.03 & 51.88 & / & 70.70 & 67.30 & / \\
\midrule
\multirow{2}{*}{LALM}
& \multirow{2}{*}{Qwen2-Audio~\cite{chu2024qwen2}} & Direct Inference & 18.96 & 39.83 & 19.84 & 53.76 & 51.52 & 47.68\\
& & CoT Inference & 26.89 & 50.57 & 28.05 & 64.33 & 60.37 & 61.61\\
\midrule

\multirow{5}{*}{LALM-FT}
& \multirow{5}{*}{Qwen2-Audio~\cite{chu2024qwen2})} 
& SFT + IR & 33.26 & 57.39 & 35.77 & 85.70 & 83.87 & 84.53\\
& & GRPO + IR & 31.60 & 55.41 & 33.22 & 81.74 & 80.00 & 80.71\\
& & + ESWR + IR & 36.23 & 63.85 & 38.57 & 84.12 & 83.90 & 83.11\\
& & + ESWR + EUR & 37.81 & 66.17 & 39.19 & 85.97 &84.85	&85.50\\
& & + ESWR + ESR & \textbf{39.46} & \textbf{69.56} & \textbf{41.87} &  \textbf{87.42}& \textbf{87.28} & \textbf{87.40}\\
\midrule
\multirow{2}{*}{/}  & \multirow{2}{*}{Comparason} 
& Ours VS SOTA & \textcolor{teal}{$\uparrow$25.1} & \textcolor{teal}{$\uparrow$34.0} & \textcolor{teal}{$\uparrow$27.1} &  \textcolor{teal}{$\uparrow$18.9} & \textcolor{teal}{$\uparrow$19.8} & \textcolor{teal}{$\uparrow$19.6} \\
& &Ours VS Baseline & \textcolor{teal}{$\uparrow$24.9} & \textcolor{teal}{$\uparrow$25.5} & \textcolor{teal}{$\uparrow$26.0} &  \textcolor{teal}{$\uparrow$6.95} & \textcolor{teal}{$\uparrow$10.9} & \textcolor{teal}{$\uparrow$10.8} \\

\bottomrule
\end{tabularx}
\end{table*}

\section{Results}
\subsection{Main Performance}
As shown in Table~\ref{tab:performance}, the results demonstrate the effectiveness of COT in LALM, our proposed ESWR, and ESR. 
\textbf{Effectiveness of COT in LALM}: Using CoT prompts significantly enhances the zero-shot SER performance of LALMs. In fact, CoT enables Qwen2-Audio to approach the performance of the best W/o-LALM pre-trained audio models without any task-specific post-training.
\textbf{Effectiveness of ESWR}: When training and testing on the same dataset, the direct use of GRPO achieves similar accuracy to SFT. This may be due to the MELD dataset containing considerable noise, resulting in the model's lower ability to recognize correct emotions. This leads to GRPO's binary rewards being too sparse, with 60\% of accuracy rewards being 0, making it difficult for the model's update policy to stabilize. However, ESWR provides more dense and psychologically grounded reward signals, improving the model's emotional reasoning capability by consistently guiding it toward the correct emotional direction. 

\textbf{The effectiveness of ESR training strategies}. Besides the ESWR method, compared to models trained with IR, models trained with EUR and ESR can both enhance emotional reasoning capabilities, improving accuracy on the MELD and IEMOCAP test sets. Moreover, models with structured thinking capabilities achieve superior accuracy relative to models lacking structured reasoning mechanisms, indicating that structured reasoning helps models avoid errors.
Through the above experiments, we have demonstrated that using the EMO-RL algorithm can significantly enhance the emotional reasoning capabilities of LALMs, achieving SOTA performance in SER tasks. Additionally, we found that our method yields greater improvements on datasets with more complex emotion labels, for example, the improvement of MELD, compared to IEMOCAP.

\begin{table}[t]
\caption{Weighted Accuracy (WA, \%) across RAVDESS, SAVEE, and IEMOCAP Datasets. The model was trained on the MELD training dataset. The baseline here denotes the SFT+IR.}
\label{tab:multi-dataset}
\centering
\small
\setlength{\tabcolsep}{4pt}
\begin{tabularx}{0.49\textwidth}{llccc}
\toprule
\textbf{Model} & \textbf{RAVDESS} & \textbf{SAVEE} & \textbf{IEMOCAP} \\
\midrule
\multicolumn{5}{l}{\textit{W/o-LALM Baselines}} \\
HuBERT large           & 25.02 & 31.54 & 44.60 \\
WavLM large            & 33.90  & 34.10 & 48.59 \\
data2vec 2.0 large     & 34.21 & 37.79 & 47.43 \\
Whisper large v3       & 40.68 & 42.18 & 46.14 \\
\midrule

\multicolumn{5}{l}{\textit{LALM-FT (Qwen2-Audio~\cite{chu2024qwen2})}} \\
 SFT+ IR          & 59.83    & 71.52  & 82.74  \\
 GRPO + BCR + IR        &  62.07   &  72.38  & 82.66 \\ 
 GRPO + ESWR + IR       & 66.21    & 74.69 & 83.05   \\
 GRPO + ESWR + EUR    &  70.43   &  78.57  & 86.11 \\
 GRPO + ESWR + ESR     & \textbf{73.99}    &  \textbf{80.83} & \textbf{87.86}  \\
\midrule
 Ours VS Baseline  & \textcolor{teal}{$\uparrow$23.67}  & \textcolor{teal}{$\uparrow$13.02}  & \textcolor{teal}{$\uparrow$6.19} \\
\bottomrule 
\vspace{-0.5cm}
\end{tabularx}
\end{table}

\subsection{Generalizability}
In practical scenarios, a model's ability to generalize emotion recognition to unseen individuals and unknown recording conditions is of paramount importance. 
To evaluate this capability, cross-dataset zero-shot testing offers an effective means of assessing a model's generalization in emotion recognition. 
We meticulously selected three diverse datasets: IEMOCAP, RAVDESS, and SAVEE. 
These datasets encompass a variety of sources, accents, and recording environments, enabling a comprehensive evaluation of the model's generalization and robustness across real-world scenarios.

As shown in Table~\ref{tab:multi-dataset}, the results of cross-datasets evaluation demonstrate that \textbf{(1)} Reinforcement learning methods, including GRPO and ESWR, demonstrate superior generalization capabilities compared to SFT methods. Notably, ESWR exhibits better generalization than GRPO. Additionally,
\textbf{(2)} Explicit Reasoning strategies show enhanced generalization over Implicit Reasoning, and Structured Reasoning strategies outperform their unstructured counterparts.
In conclusion, the combination of ESWR and ESR surpasses all baseline and alternative training methods, achieving the highest performance in emotional reasoning generalization.

\begin{table}[t]
\caption{Performance of quantitative ablation of the reward mechanism alone on MELD dataset. The model was trained based on Qwen2-Audio}
\label{tab:ablation}
\centering
\small
\begin{tabularx}{0.48\textwidth}{Xccc}
\toprule
\textbf{Method } & \textbf{UA(\%)} & \textbf{WA(\%)} & \textbf{F1(\%)} \\
\midrule
GRPO+IR & 18.22 & 38.32 & 18.96 \\
GRPO+EUR & 25.55 & 49.12 & 26.36 \\
GRPO+ESR & 29.19 & 53.53 & 30.61 \\
GRPO+BCR & 33.62 & 55.42 & 35.57 \\
GRPO+ESWR & 35.02 & 62.93 & 37.73 \\
\bottomrule 
\end{tabularx}
\end{table}

\begin{figure}[t]
  \centering
\includegraphics[width=0.49\textwidth]{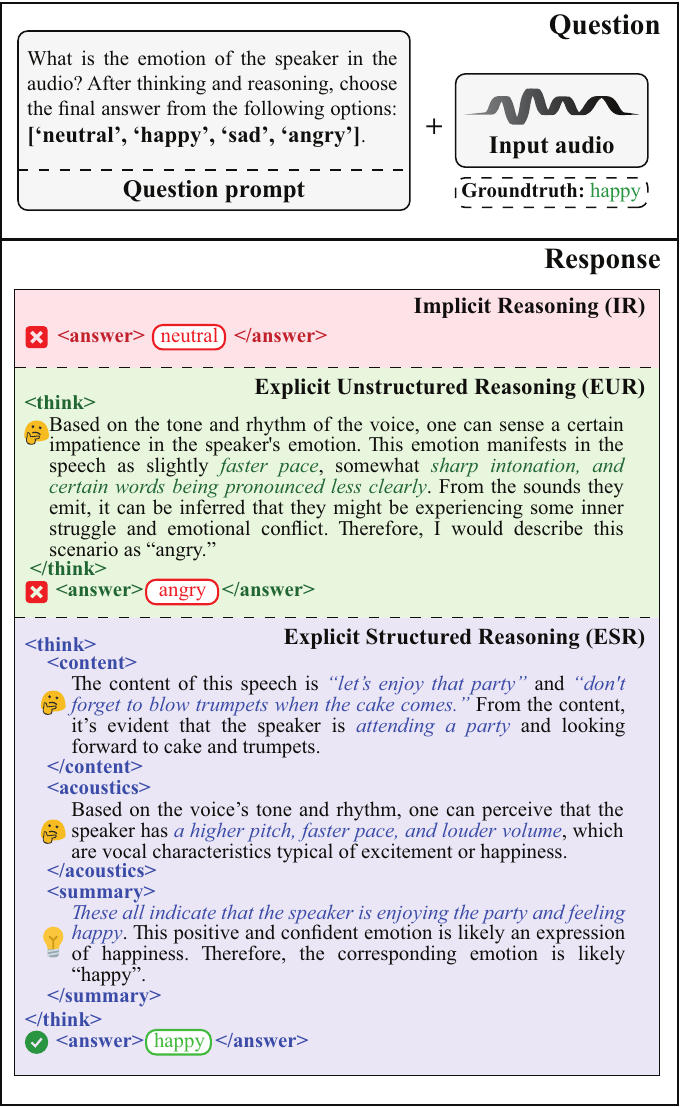}
  \caption{An example of the reasoning results of IR, EUR, and ESR}
  \label{fig:case_study}
  \vspace{-0.5cm}
\end{figure}

\subsection{Ablation }
To explore the quantitative ablation of the effects of the reward mechanism alone. As shown in Table~\ref{tab:ablation}, we have supplemented the relevant quantitative ablation experiments on MELD dataset. For the methods in the first three rows, we only used the corresponding format reward without the accuracy reward. For the last two rows, we only used the corresponding accuracy reward without the format reward. These results demonstrate the efficiency of our proposed ESWR and ESR methods for speech emotion recognition.

\subsection{Case Study}
In Figure~\ref{fig:case_study}, we show a case study that demonstrates the response results when testing the same speech sample after training with different methods.
Models trained with IR seem to have lost many other abilities, such as not trying to think and reasoning, even though I asked them to do so.

Models trained with EUR can generate flexible reasoning based on different speech inputs. They often analyze emotions primarily through acoustic features such as pitch, rhythm, and speed. While this approach is effective for simpler cases, it faces challenges with more complex scenarios due to the omission of critical semantic emotional details.

In contrast, models trained with ESR explicitly document the speaker's key content and acoustic features, followed by a comprehensive analysis of both semantic and auditory information. This structured approach reduces the likelihood of overlooking key details, thereby enhancing the model's emotional reasoning capabilities.



\section{Conclusion}
This paper introduces EMO-RL, a reinforcement learning framework that improves the emotional-reasoning capacity of large audio–language models for speech-emotion recognition.
By incorporating emotion similarity-weighted reward, which integrates psychological prior knowledge into RL, and Explicit Structured Reasoning into our framework, EMO-RL effectively overcomes the challenges of convergence instability and limited reasoning ability in speech emotion recognition tasks. 
Comprehensive experiments demonstrate that EMO-RL not only improves the emotional reasoning capabilities of LALMs on the MELD and IEMOCAP datasets (compared with SOTA, achieving an UA improvement of 25.1\% and 18.9\%, respectively), but also shows excellent generalization across different datasets. 
This work signifies a step forward in applying reinforcement learning and large audio-language models to speech emotion recognition, paving the way for future speech affective computing research. Moreover, EMO-RL shows potential for enhancing multi-modal LLMs' emotion perception, bringing us closer to building truly emotional LLMs.

\section{Limitation}
Our proposed method has certain limitations that warrant attention.
Firstly, while our EMO-RL framework is designed to be versatile and applicable across a variety of multi-modal scenarios, including video, audio, and text, our current experimental scope has been limited to the speech modality alone. We have not yet incorporated visual elements such as images or videos into our experimental design. This restriction means that the full potential of our framework in multi-modal contexts remains unexplored. 
Secondly, although exploiting the LALMs for SER tasks has delivered promising results, it has also introduced challenges related to computational complexity and inference efficiency. 
The inference efficiency of our approach is comparatively lower than that of previous methods, which might affect its practicality for real-time applications. 
In the future, we will try to solve the above limitations.

\section{Ethical Considerations}
The deployment of SER systems raises significant ethical concerns that build upon established frameworks for sentiment and emotion analysis. Privacy and consent represent primary issues, as SER extracts sensitive psychological information from vocal patterns often without users' explicit awareness, unlike voluntary text-based sentiment analysis. Additionally, SER systems exhibit systematic biases across demographic groups and may misinterpret cultural differences in emotional expression, with training datasets often lacking diverse representation—problems shared with broader emotion analysis research. The "black box" nature of deep learning-based systems also raises accountability concerns when informing decisions affecting individuals' lives. These considerations highlight the need for robust consent frameworks, diverse datasets, and ethical guidelines specific to speech emotion recognition that address the unique challenges of speech emotion detection.

\section{Acknowledgements}
This work was supported by the Shenzhen-Hong Kong Joint Funding Project (Category A) under Grant No. SGDX20240115103359001.

\bibliography{custom}
\end{document}